\documentclass[10pt,letterpaper,amsmath,amssymb,final]{article}
\usepackage[top=1.in, bottom=1.in, left = 1in, right=1in]{geometry}
\usepackage{graphicx}
\title{WLS R\&D for the Detection of Noble Gas Scintillation at LBL: seeing the light from neutrinos, to dark matter, to double beta decay}

\author{Victor M. Gehman\footnote{vmgehman@lbl.gov}\\{\small Lawrence Berkeley National Laboratory, Physics Division}\\{\small Berkeley, CA 94720}}

\begin{document}

% Make the title...
\maketitle

\abstract{Radiation detectors with noble gasses as the active medium are becoming increasingly common in experimental programs searching for physics beyond the standard model.  Nearly all of these experiments rely to some degree on collecting scintillation light from noble gasses.  The VUV wavelengths associated with noble gas scintillation mean that most of these experiments use a fluorescent material to shift the direct scintillation light into the visible or near UV band.  We present an overview of the R\&D program at LBL related to noble gas detectors for neutrino physics, double beta decay, and dark matter.  This program ranges from precise measurements of the fluorescence behavior of wavelength shifting films, to the prototyping of large are VUV sensitive light guides for multi-kiloton detectors.}

%\keywords{Neutrinos; Dark Matter; Noble Gasses; Radiation Detectors; Scintillation Light; Vacuum Ultraviolet Light; Wavelength Shifters}

%\begin{document}

\section{Introduction and Overview}\label{sec:Intro}
Numerous experiments searching for answers to some of the most important and fundamental questions in physics today, center on the use of noble gases as particle detectors. This program is incredibly rich and diverse, ranging from: direct searches for Weakly Interacting Massive Particle (WIMP) dark matter\cite{XENON, XMASS, LUX, MiniCLEAN, DEAP3600, CLEAN, DARKSIDE}, general particle detectors, especially neutrinos \cite{ICARUS, MicroBooNE, MEG, EXO, NEXT, LBNE, Lanou1987, McKinsey2000}, neutron electric dipole moment (EDM) searches\cite{nEDM1, nEDM2}, and neutron lifetime measurements\cite{Huffman2000}.  

Most of the detector systems listed above detect the scintillation light from particle interactions in the noble gas because noble gasses have remarkably high scintillation yield (typically between 20 and 40 photons per keV\cite{Lippincott2010}).  The principle difficulty associated with noble gas scintillation is that the photons are emitted in the vacuum ultraviolet (VUV), with wavelengths ranging from approximately 80 nm (for helium and neon) to 128 nm (for argon) and 175 nm (for xenon).  Photons in this band are difficult to detect because they are short enough in wavelength to scatter on most chemical bonds (and thus are strongly absorbed by most materials, even those normally used for optical windows), but are not high enough in energy to be treated calorimetrically (as one would do with x rays or $\gamma$ rays).

Most detectors using noble gas scintillation light exploit some fluorescent material to shift the VUV photons into the visible band (note that because xenon scintillates around 175 nm, where quartz is still transparent, some xenon-based detectors are designed to be directly sensitive primary scintillation.  Noble gasses are also nearly perfectly transparent to their scintillation photons, so these fluorescent materials do not have to be dissolved or suspended in the bulk detector.  Instead, they tend to be films deposited onto surfaces around the edge of the noble gas volume.  Most of these films are organic molecules containing one or more phenyl groups.  A popular and well-characterized wavelength shifting (WLS) material is ``1,1,4,4-Tetraphenyl Butadiene'' (TPB)\cite{McKinsey1997, Jones2012, Gehman2011}. In particular, TPB has been used in several experiments that rely on detecting helium scintillation light\cite{Huffman2000,McKinsey2004,Archibald2006,Ito2012}.

\section{Experimental Hardware and Measurement Strategy}\label{sec:AbsEff}
The experimental apparatus used in the studies outlined in this article is built around a McPhereson model 632 deuterium lamp and a model 234/302VM monochromator. The entire optical train is under vacuum (P $\sim\ 10^{-4}$ mBar). The WLS plastic samples are deposited on one-inch diameter discs made from Solacryl SUVT acrylic from Spartech Polycast\cite{SolacrylDatasheet}.  Sample discs are mounted in a vacuum manipulable filter wheel so that different samples, lamp measurements, and dark current could all be measured quickly to avoid any time drift in the system over more than a few minutes.  Light detection was done with one of two photon sensors: fluorescence spectral shapes were measured with an Ocean Optics QE65000 UV/Vis spectrometer (sensitivity from 200 to 1000 nm), and absolute intensity measurements were made with a calibrated photodiode from International Radiation Devices.  Details of the hardware and analysis for this and similar measurements can be found in Reference \cite{Gehman2011}.

\section{Recent Results}\label{sec:Results}
We now present an overview of recent results in the measurement of fluorescence behavior of several WLS materials.  These measurements were made around the same time as those for \cite{Gehman2013}, but have yet to be published in an article of their own.

\subsection{Fluors Other than TPB}\label{sec:OtherFluors}
\begin{figure}[h]
\begin{center}
\includegraphics[angle=0, width=10cm]{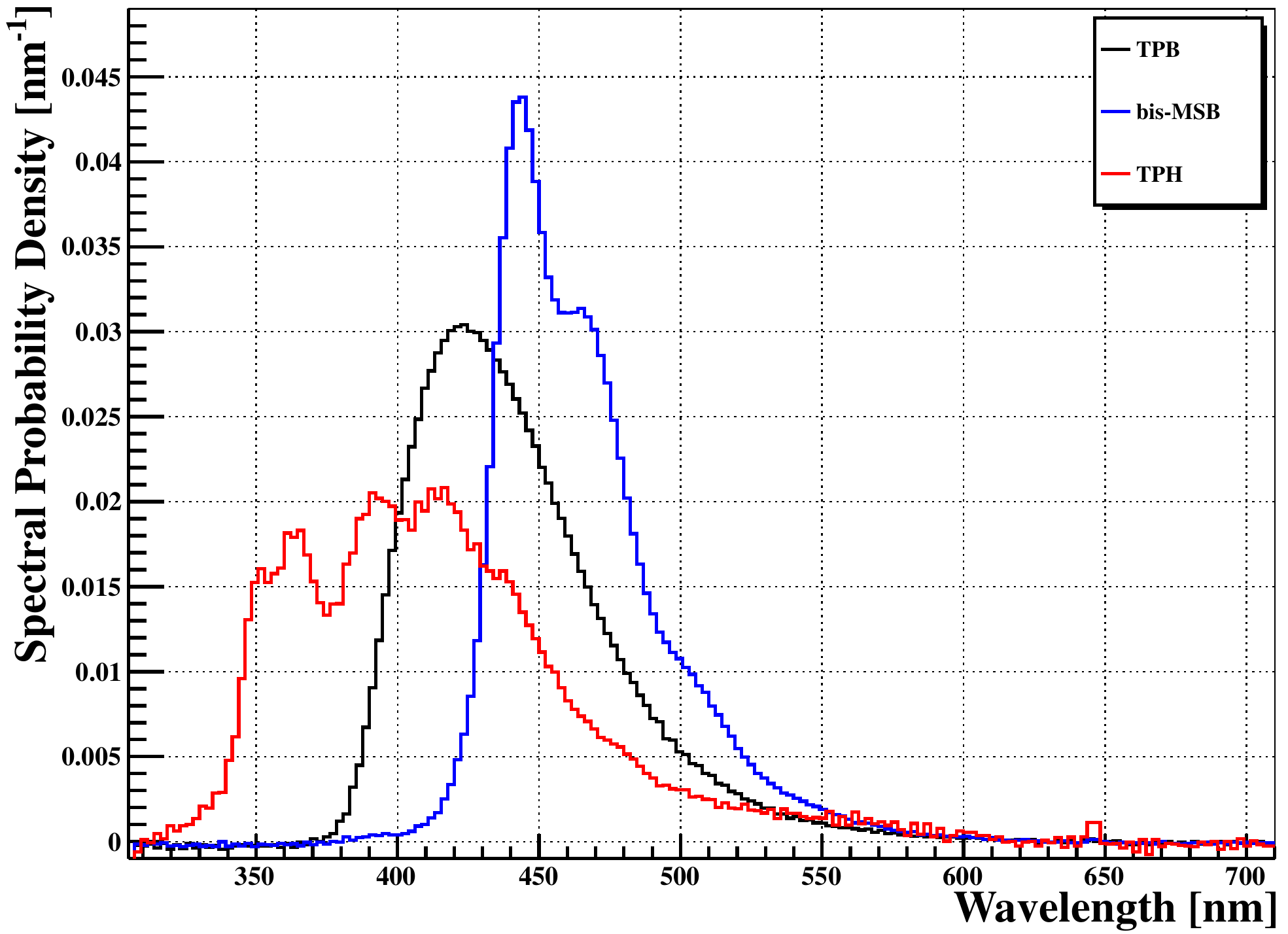}
\caption{Fluorescence spectra for TPB (black), bis-MSB (blue), and TPH (red) as a function of wavelength.  All three spectra have been normalized to unit area to facilitate shape comparison.}
\label{fig:OtherWLSFlSpec}
\end{center}
\end{figure}

\begin{figure}[h]
\begin{center}
\includegraphics[angle=0, width=10cm]{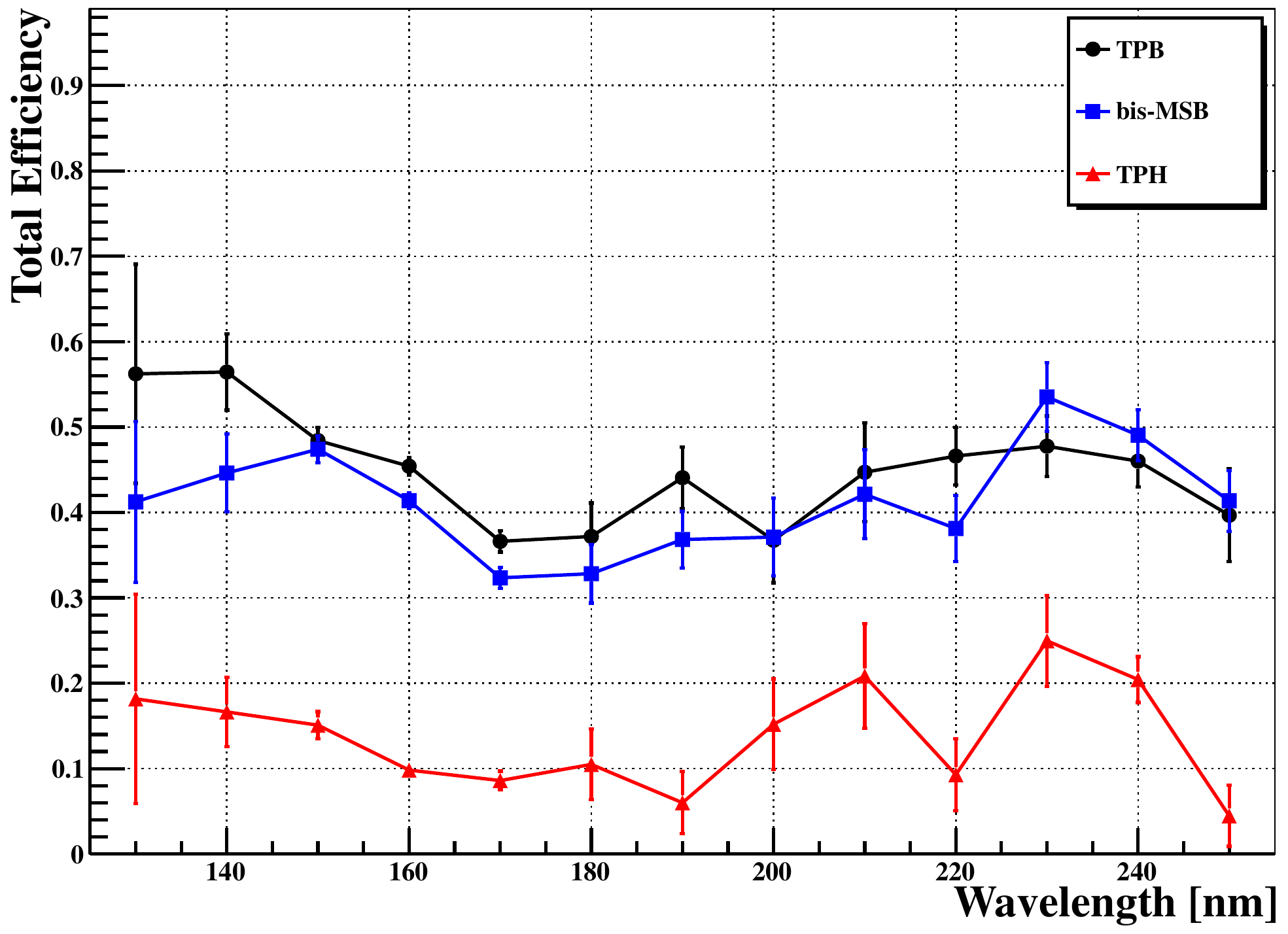}
\caption{Total fluorescence efficiency for TPB (black), bis-MSB (blue), and TPH (red) as a function of wavelength.}
\label{fig:OtherWLSEff}
\end{center}
\end{figure}

We embarked on a campaign to study the fluorescence behavior of wavelength shifting plastics other than TPB.  The LBNE experiment has been investigating the use of 1,4-Bis (2-methylstyryl) benzene (bis-MSB) as a wavelength shifter to convert liquid argon scintillation to visible light (there is quite a lot of experience in the No$\nu$A experiment using bis-MSB in organic liquid scintillator).  p-Terphenyl (TPH) is another wavelength shifter commonly used in organic liquid scintillator.  We examined the fluorescence spectrum and conversion efficiency for both of these fluors.  Figure \ref{fig:OtherWLSFlSpec} shows the fluorescence spectra for both of these fluors as well as that for TPB to serve as a reference.
Once we measured the fluorescence spectrum for each of the new fluors, we could then measure the total fluorescence efficiency.  Efficiency curves for bis-MSB and TPH (again along with TPB for comparison) in Figure \ref{fig:OtherWLSEff}.
Closer examination of Figure \ref{fig:OtherWLSFlSpec} shows that these different fluors have rather different fluorescence spectra.  These data and further studies of other WLS materials allow for the engineering of the optical systems for the collection of VUV photons, where the emission spectrum of the fluor can be carefully matched to the sensitivity of visible photon collectors.  Clearly, future fluor characterization measurements will contribute to this capability in the community as well.  Figure \ref{fig:OtherWLSEff}, shows us that bis-MSB is indeed a viable alternative to TPB in terms of its fluorescence efficiency, and should be seriously considered in situations where cost, durability, or optical properties favor it.

\subsection{R\&D for Large-Area Collectors}\label{sec:GoBig}
As detector masses increase for current and future experiments, the total area of photon collection surfaces must also increase if reasonable light collection efficiencies are to be maintained.  For this reason we have begun to examine a number of options for large area VUV sensitive surfaces.  First, and most directly related to LBNE, we have investigated the use of large WLS-doped plastic panels in cooperation with Eljen Technology\cite{Eljen}.  To this end, we purchased several 1.5-inch square tiles, doped with three different concentrations of two different wavelength shifters (TPB and bis-MSB), and interrogated them with the VUV light source shown above, reading out the fluorescence yield with the same photodiode used to make the total fluorescence efficiency measurements discussed in this article.
\begin{figure}[h]
\begin{center}
\includegraphics[angle=0, width=10cm]{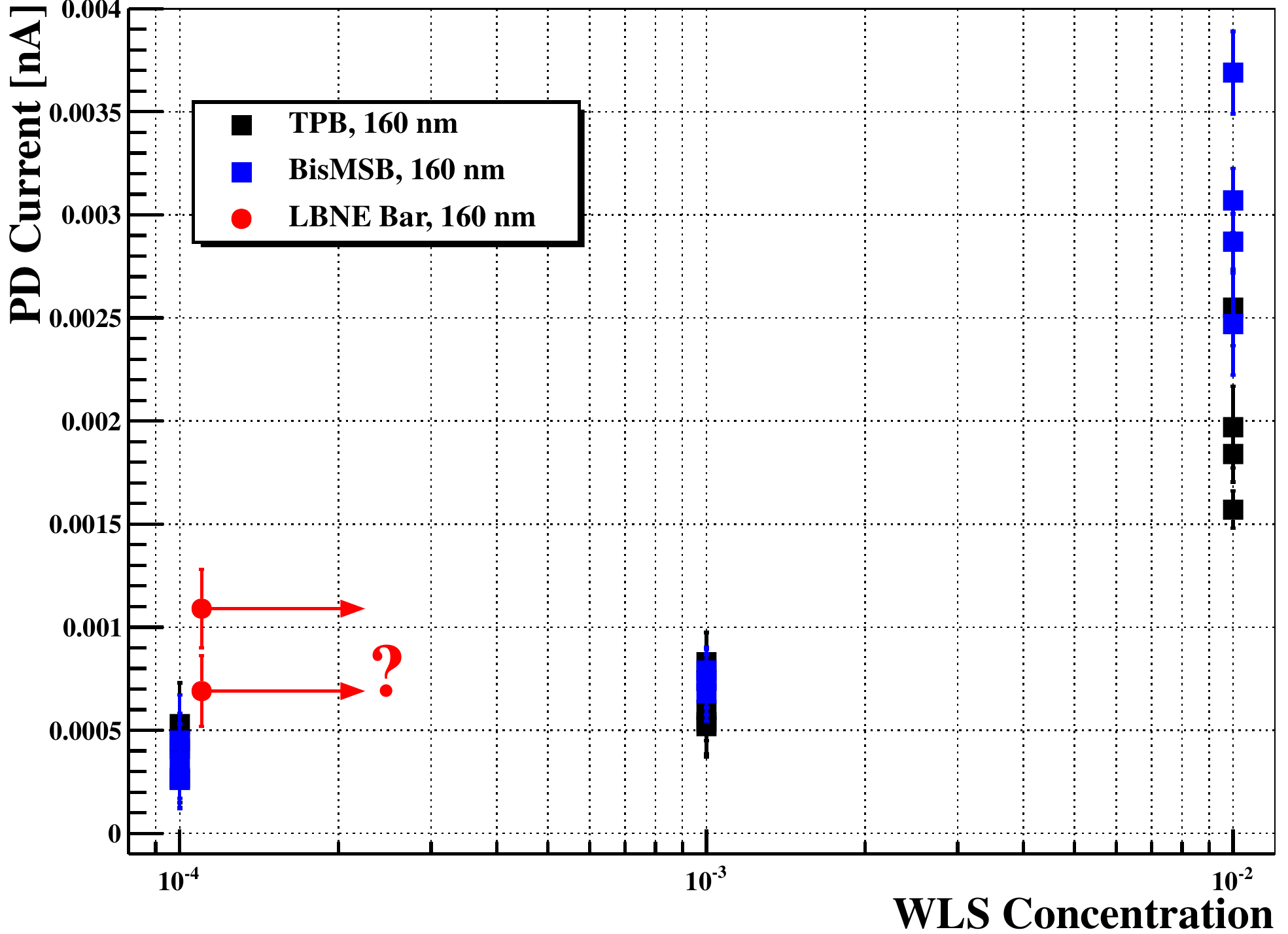}
\caption{Relative fluorescence light yield as measured by photodiode current for three concentrations (0.01\%, 0.1\%, and 1\%) for two wavelength shifters (TPB in black and bis-MSB in blue) embedded in polystyrene tiles.  For comparison, the baseline WLS bars for LBNE are also plotted in red.}
\label{fig:WLSTiles}
\end{center}
\end{figure}
In Figure \ref{fig:WLSTiles}, we also plot the fluorescence yield from a 1.5 inch long segment of one of the WLS bars which are currently the LBNE baseline.  Because the LBNE baseline bars are coated rather than doped with their fluor, it is not clear where its points should be plotted on the horizontal axis, and we have attempted to indicate that in this figure.  The data in Figure \ref{fig:WLSTiles} clearly show that more WLS dramatically improves fluorescence light yield, and we are preparing for a physically longer test sample that will be a one-for-one "drop-in replacement" for the LBNE baseline bars.  These longer bars will probably be cast into PMMA or PEMA acrylic rather than polystyrene because latter has a tendency to craze at low temperature.  If the optical attenuation is good enough in these WLS-doped bars, we would ideally like to build very large WLS doped panels that would fill most or all of the space inside the space inside the LBNE anode plane assemblies\cite{LBNE}.

\begin{figure}[h]
\begin{center}
\includegraphics[angle=0, width=10cm]{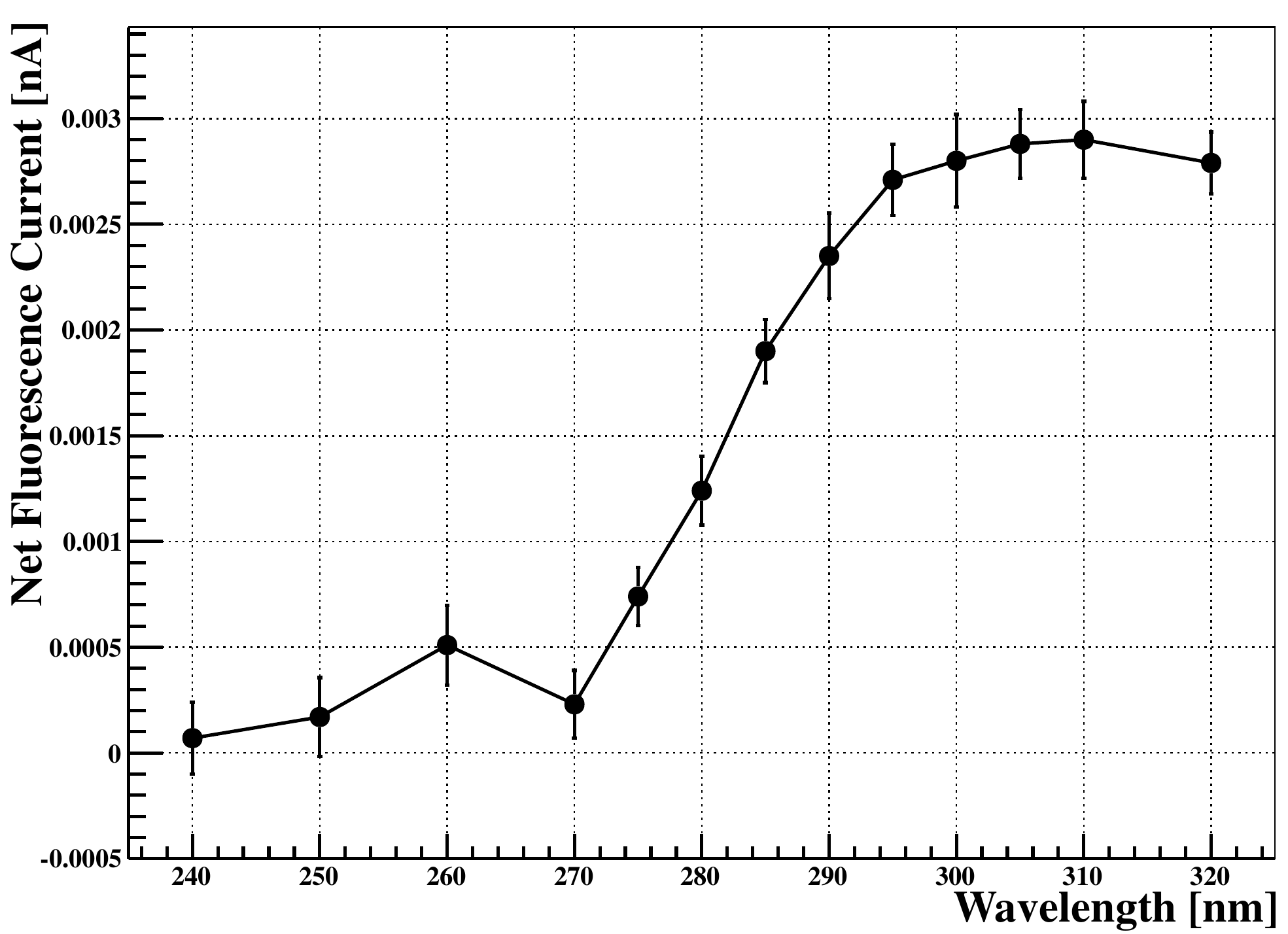}
\caption{Relative fluorescence light yield measured by photodiode current for EJ-299-15 doped PMMA bar.}
\label{fig:WLSBar}
\end{center}
\end{figure}

We are investigating another photon collection device for an experiment that uses a wavelength shifting gaseous molecule mixed into a high-pressure xenon time projection chamber.  This WLS gas, Trimethylamine (TMA), absorbed the 175-nm light from xenon, and fluoresces at roughly 300 nm.  The goal here is to collect that secondary fluorescence light with WLS bars around the edge of the TPC volume.  We are investigating a number of commercial options, and have already begun preliminary testing on one of them (one-inch by half-inch by twelve-inch PMMA bar doped with EJ-299-15, also from Eljen).  In Figure \ref{fig:WLSBar}, we see the relative fluorescence yield as a function of input VUV wavelength.
We have another commercial option from Eljen to test soon and we will expand this program to investigate a number of other options in the near future.  One last note is that the fluorescence yields discussed here in section \ref{sec:GoBig} are relative measurements to inform our understanding of the shape of the fluorescence response.  To convert this into an absolute fluorescence photon yield we have to carefully survey the inside of our vacuum chamber in the configuration required to study the fluorescence behavior of these samples that are too large to fit in the filter wheel, and calculate the geometric corrections to the photon collection efficiency.  We are beginning this process now, but have not yet completed it.

\section{Conclusions}\label{sec:Concl}
Noble gas based radiation detectors have been an important component of the search for physics beyond the Standard Model and will continue to increase the size of that role in the coming years.  An important part of this process is carefully understanding the efficiency with which we can collect scintillation photons from these detectors.  Just as important is the process of carefully designing the photon detection train, in terms of matching up absorption and emission spectra with detector sensitivities to maximize this photon detection efficiency.  In this article, we have outlined a program to begin to address these issues.  We have already answered a number of questions, and are in the process of continuing to upgrade and better characterize our experimental hardware, to continue in this endeavor.

\section{Acknowledgments}
This work was supported by the Director, Office of Science, of the U.S. Department of Energy under Contract No. DE-AC02-05CH11231.

\end{document}